\documentclass[twocolumn,secnumarabic,amssymb, nobibnotes, aps, prc,nofootinbib,fleqn, floatfix,showpacs,superscriptaddress]{revtex4-1}

\usepackage{graphicx}
\usepackage{dcolumn}
\usepackage{mathrsfs}
\usepackage{bm}
\usepackage{dsfont}
\usepackage{graphicx}
\usepackage{dcolumn}
\usepackage[usenames]{color}
\begin{document}

\title{Exact solution of the pairing problem for spherical and deformed systems}
\author{Chong Qi}
\email{chongq@kth.se} 
\affiliation{Department of Physics, Royal Institute of Technology (KTH), SE-10691 Stockholm, Sweden}
\author{Tao Chen}
\affiliation{Department of Physics, Stockholm University, Stockholm, SE-106 91, Sweden}

\begin{abstract} 
There has been increasing interest in studying the Richardson model from which one can derive the exact solution for certain pairing Hamiltonians. However, it is still a numerical challenge to solve the nonlinear equations involved. In this paper we tackle this problem by employing a simple hybrid polynomial approach. The method is found to be robust and is valid for both deformed and nearly spherical nuclei. It also provides important and convenient initial guesses for spherical systems with large degeneracy. As an example, we apply the method to study the shape coexistence in neutron-rich Ni isotopes.
\end{abstract}

\maketitle
The pairing correlation has long been recognized as an important residual correlation in atomic nuclei and other many-body systems including superconductors, neutron stars and trapped two-component Fermi gases. The simple Bardeen-Cooper-Schrieffer (BCS) approximation and the generalized Hartree-Fock-Bogoliubov theory are successful in describing the pairing properties of open-shell nuclei. The drawbacks of such approaches have also been known, which, in particular, include the particle number fluctuations and the collapsed condensate around nuclear shell closures. Extensive efforts have thus been done in developing  alternative pairing models in order to overcome those drawbacks, which include the exact (and Lanczos) diagonalization and the Richardson (or the Richardson-Gaudin) method (see, Ref. \cite{Chen2014} and references therein and recent reviews in Refs. \cite{Isa14,Duk04}).
Within the Richardson model \cite{Rich63}, a system with a constant pairing coupling satisfies a set of nonlinear Richardson (or Bethe ansatz) equations as
\begin{eqnarray}\label{rich}
1+\sum_j\frac{\Omega_jG}{x_i-2\varepsilon_j}-\sum_{k\neq i}\frac{2G}{x_i-x_k}=0,
\end{eqnarray}
where $G$ is the strength of the pairing coupling, $\varepsilon_j$ denotes the single-particle energy of the orbital $j$ with degeneracy $\Omega_j$, $x_i$ are the Richardson variables (or pair energies).
The total energy is given as a sum of Richardson variables as $E=\sum^N_ix_i$. The advantages of the above model over the exact diagonalization method include: There is no dimension limitation and the Richardson variable thus defined may  provide interesting information on the pairing correlation of each pair.
The latter problem can be of fundamental importance by considering the fact that all pairs condense to the same state within the BCS ansatz wheres they are all different within the exact model.

It is recognized that the nonlinear coupled equations as defined in Eq. (\ref{rich}) are very difficult to solve since they contain singularities and can become numerically unstable, that is, when one certain variables $x_i$ approach twice the single-particle energies and coincide with other variables. On top of that, the variables $x_i$ can take either real or complex values.
For examples, for two pairs in a single-$j$ shell, $x'$ are complex numbers irrespective of the coupling strength
\begin{eqnarray}
x_{1,2}= (1-\Omega_j)G \pm\sqrt{\Omega_j-1}Gi.
\end{eqnarray}
For two pairs in two doubly-degenerate orbitals separated by $d$, one has
\begin{eqnarray}
x_{1,2}=2\varepsilon_{1,2}-G \pm[d-\sqrt{d^2-G^2}],
\end{eqnarray}
which become a complex conjugate pair with $G>d$.
 Moreover, there are as many independent solutions as seniority  $v=0$ states contained in the model space. The solutions are sensitive to the initial values,  which makes it difficult to locate the desired state.  There have been many efforts trying to tackle these problems. In particular, the polynomial approaches as defined in Refs. \cite{Far11, Guan12} look quite promising. In this paper we are interested in exploring further 
in that direction. Our aim is to find a robust and practical approach that can be conveniently applied in solving the nuclear pairing problem.

Firstly we go through the basic ideas of the polynomial approach in a way that is slightly different from Refs. \cite{Far11,Guan12} but may be easier to understand. Following Refs. \cite{Far11,Guan12}, one defines a polynomial of degree $N$ as
\begin{equation}\label{poly}
P(x)=\prod_{i=1}^N (x-x_i) =\sum_j^N a_jx^j,
\end{equation}
where $N$ is the number of pairs and Richardson variables and $a$ the expansion coefficients. In this approach, $a$ become the unknown variables instead of the Richardson variables $x_i$. The $N$  roots of above polynomial correspond to the values of the Richardson variables. In particular, we have
\begin{equation}
a_1  =-\sum_i^N x_i= -E.
\end{equation}
That is, it corresponds to the opposite value of the total energy of the system. The advantages of the polynomial approach include: The coefficients $a$ are all real and one avoids finding the roots of the polynomial if only the binding energy is of interest. 

The first and second derivatives of the polynomial satisfy the relation
\begin{equation}
\frac{P'(x)}{ P(x)}=\sum_{i=1}^N \frac{1}{x-x_i},
\end{equation}
and
\begin{eqnarray}\label{poly2}
\nonumber \frac{P''(x)}{ P(x)}&=&\left(\frac{P'(x)}{ P(x)}\right)^2+\left(\frac{P'(x)}{ P(x)}\right)'\\
\nonumber &=& \sum_{i,j=1}^N \frac{1}{(x-x_i)(x-x_j)}-\sum_{i=1}^N \frac{1}{(x-x_i)^2}\\
\nonumber &=& \sum_{i\neq j}^N \frac{1}{(x-x_i)(x-x_j)}\\
&=& \sum_{i\neq j}^N \frac{2}{(x-x_i)(x_i-x_j)},
\end{eqnarray}
where one applied the simple relation that $1/(x-x_i)(x_i-x_j)+1/(x-x_j)(x_j-x_i)=1/(x-x_i)(x-x_j)$.
Combining Eqs. (\ref{rich}) and (\ref{poly2}), we have,
\begin{eqnarray}\label{poly21}
\nonumber \left(\frac{P'(x)}{ P(x)}\right)^2+\left(\frac{P'(x)}{ P(x)}\right)'\\ 
\nonumber= \sum_{i}^N \frac{1}{G(x-x_i)}
-\sum_{i=1}^N\sum_{j=1}^M \frac{\Omega_j}{(x-x_i)(2\varepsilon_j-x_i)}\\
\nonumber= \frac{1}{G}\left(\frac{P'(x)}{ P(x)}\right)\\
+\sum_{j=1}^M \frac{\Omega_j }{x-2\varepsilon_j}\left[\left(\frac{P'(x)}{ P(x)}\right)-\left(\frac{P'(2\varepsilon_j)}{ P(2\varepsilon_j)}\right) \right],
\end{eqnarray}
where $M$ denotes the total number of single particle orbitals.
Above expressions are valid for any $x$. One interesting observation is that, if the value of $x$ approaches twice the single-particle energy of a given orbital $j_{\delta}$, i.e., $x=2\varepsilon_{\delta}$, one has
\begin{eqnarray} \label{poly3}
\nonumber \left(\frac{P'(x_{\delta})}{ P(x_{\delta})}\right)^2+ (1-\Omega_{\delta}) \left(\frac{P'(x_{\delta})}{ P(x_{\delta})}\right)' - \frac{1}{G}\left(\frac{P'(x_{\delta})}{ P(x_{\delta})}\right)\\
=\sum_{j\neq\delta}  \frac{\Omega_j }{2\varepsilon_{\delta}-2\varepsilon_j}\left[\left(\frac{P'(2\varepsilon_{\delta})}{ P(2\varepsilon_{\delta})}\right)-\left(\frac{P'(2\varepsilon_j)}{ P(2\varepsilon_j)}\right) \right]  
\end{eqnarray}

As shown in Ref. \cite{Far11}, if all orbitals are doubly degenerate, above equation set reduces to a much simpler one as
\begin{eqnarray}\label{poly4}
\nonumber \left(\frac{P'(x_{\delta})}{ P(x_{\delta})}\right)^2 - \frac{1}{G}\left(\frac{P'(x_{\delta})}{ P(x_{\delta})}\right)\\
-\sum_{j\neq\delta}  \frac{\left[\left(\frac{P'(2\varepsilon_{\delta})}{ P(2\varepsilon_{\delta})}\right)-\left(\frac{P'(2\varepsilon_j)}{ P(2\varepsilon_j)}\right) \right] }{2\varepsilon_{\delta}-2\varepsilon_j} =0.
\end{eqnarray}
The basic idea of Ref. \cite{Far11} is to firstly find the values for the log-derivatives ${P'(x_{\delta})}/{ P(x_{\delta})}$ by solving above equation and, in a second step, determine $a$ values from those log-derivatives. However, in that case we will have more unknown variables than the original Eq. (\ref{rich}) for partially occupied systems since one has $M>N$.

Alternatively one can solve Eq. (\ref{poly3})  directly by inserting the expression for the polynomial Eq. (\ref{poly}) and choosing $N$ number of single-particle orbitals $j_{\delta}$, which will be well defined as long as the orbitals chosen are doubly degenerate with $\Omega_{\delta}=1$. One does not require all orbitals to be doubly degenerate. This approach is particularly convenient if only the ground state and low-lying states are of interest. In that case one can start from the lowest-lying single-particle orbitals within the well defined Hartree-Fock (HF) configuration. As for the ground state, the HF energy also provides an clear upper bound for $-a_1$.

Eqs. (\ref{poly3}) and (\ref{poly4}) are enough for most nuclear pairing calculations by taking into account the fact that many nuclei are deformed or weakly deformed (see, e.g., recent calculations in Refs. \cite{Wu15,Cha15a} and references contained therein), which can be well described within the doubly degenerate Nilsson scheme.
However, the equation set becomes ill defined or instable if the system is spherical with larger degeneracy or nearly spherical.
The possible generalization of Eq. (\ref{poly4}) to degenerate systems is discussed in Ref. \cite{Far11}. Here we turn to a possibly simpler approach.
We give a simple derivation for the polynomial approach as applied in Ref. \cite{Guan12} by rewritting Eqs. (\ref{poly2}-\ref{poly21}) as
\begin{eqnarray}
\nonumber P''(x)- \left(\frac{1}{G}+\sum_{j}  \frac{\Omega_j }{x-2\varepsilon_j} \right) P'(x)\\
+ \sum_{j=1}^M \frac{\Omega_j }{x-2\varepsilon_j} \left(\frac{P'(2\varepsilon_j)}{ P(2\varepsilon_j)}\right) P(x)=0. 
\end{eqnarray}
This equation can be reexpressed as a polynomial of degree $N+M-1$
\begin{eqnarray}\label{poly5}
A(x)P''(x)- B(x) P'(x)
+ C(P,x) P(x)=0 
\end{eqnarray}
where $A(x)=\prod_{j=1}^M (x-2\varepsilon_j)$, $B(x)=  \left(\frac{1}{G}+\sum_{j}  \frac{\Omega_j }{x-2\varepsilon_j} \right)A(x)$ and 
$C(P,x) = \sum_{j=1}^M \frac{\Omega_j }{x-2\varepsilon_j} \left(\frac{P'(2\varepsilon_j)}{ P(2\varepsilon_j)}\right)A(x)$ can all be expressed as polynomials of the degrees $M$, $M-1$ and $M-1$, respectively. The unknown variables are again only the expansion coefficients $a_i$. They can be determined by applying the condition that above equation is valid for any $x$ and, therefore, all coefficients at different orders $x^j$ must be zero. The polynomial $C$ is, however, not defined in Ref. \cite{Guan12}. The relation of $C$ to the polynomial $P$ in the large $G$ limit was given in Eqs.(19-21) in Ref. \cite{Pan13}. In Ref. \cite{Guan12} $C$ was derived from above equation through symbolic calculations which can be very time consuming. On top of that, one has to deal with as many as $N+M$ polynomials. Eq. (\ref{poly5}) is valid for both spherical and deformed systems and is numerically stable to solve. The drawback is that it contains trivial non-solutions with $A(x)=0$, where the Richardson variables coincide with twice the energy of any of the single-particle orbitals, $x_i=2\varepsilon_j$. Those trivial non-solutions may be difficult to avoid if one does not have a good initial guess.

We have developed several Python codes to solve Eqs. (\ref{poly3}), (\ref{poly4}), (\ref{poly5}) following the discussions described above as well as  the method as described in Ref. \cite{Guan12}. Our calculations show that all those codes work well. They will be available to the public after further testing and documentating. We did not explore the full capacity and computation limitation of the present approaches. The largest problems we handled at the current stage is a half-filled system with 25 pairs and 50 orbitals, which has a dimension $1.26*10^{14}$ and is practically enough for all nuclear pairing problems. 
It takes rough 0.3, 5 and 170 seconds to solve half-filled systems with 10, 20 and 50 orbitals, respectively, on a usual desktop with Intel Core i7-2600k 3.40GHZ processor (single thread). The most time consuming part corresponds to the construct of the equation sets, which can be parallelized in a straightforward way. 
It may be useful to mention that we specifically choose the Python language by considering the fact that it can be easily connected to many well-established nuclear theory codes which are mostly done in Fortran. Calculations are also compared with those given by our exact diagonalization approach \cite{Xu2013247,Cha15}.

In the following we apply the algorithms developed to study the ground state of different systems.
The correlation energy of the ground state relative to the HF configuration can then be given as
\begin{eqnarray}
 E^{(1)}_{\rm corr} =\sum^N_ix_i-(2\varepsilon_i-G),
\end{eqnarray}
where the Richardson variables are ordered according to their energies and $\varepsilon_i$ corresponds to the single-particle energy of the $i$-th pair within the HF configuration. It is possible to define the correlation energy gain by each pair within the Richardson model
\begin{equation}
x'_i=x_i-2\varepsilon_i+G,
\end{equation}
where both Richardson variables and the single-particle energies are ordered according to their energies.

\begin{figure}  
\begin{center}
\includegraphics[width=0.45\textwidth]{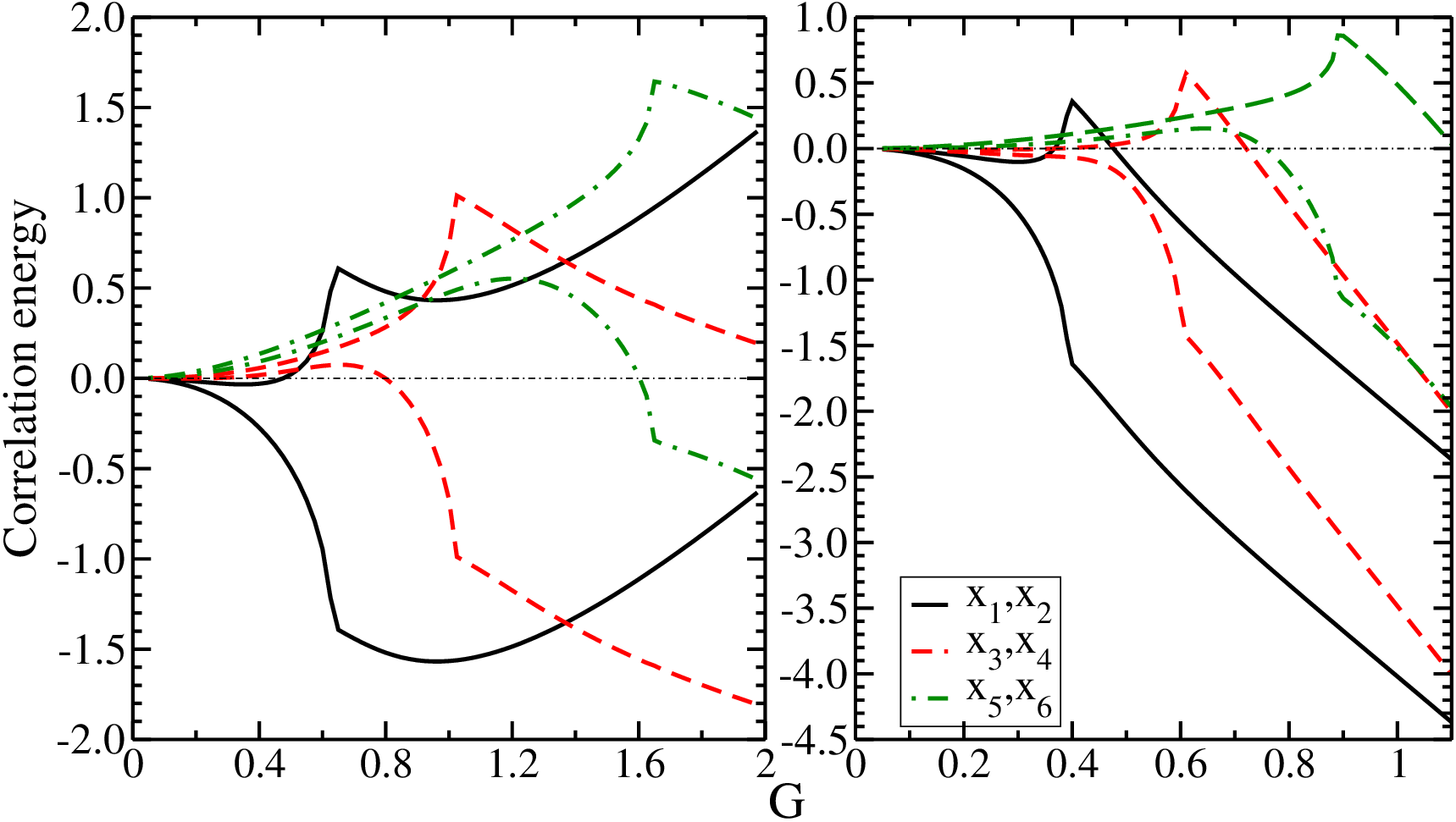}
\end{center}
\caption{\label{fig:Tin161_j} (color online) Correlation energies for each pair, $x'_i=x_i-2\varepsilon_i+G$, as a function of the pairing strength, $G$, within the Richardson model for a fully-occupied system (left) and a half-occupied (right) system with $N=6$ pairs and, respectively, 6 and 12 equally-spaced doubly-degenerate orbitals separated by one unit. The kinks corresponds to where the transition from real to complex numbers of the Richardson variables occurs. In the left case, the summation over all $x'_i$ equals zero.
}
\end{figure}

As an example, in the right panel of Fig. \ref{rich} we plot the correlation energy gain by each pair, $x'_i=x_i-2\varepsilon_i+G$, for a half-occupied system with $N=6$ pairs in 12 equally-spaced doubly-degenerate orbitals. The orbitals are separated by one unit. Exact solutions for this chosen example was known \cite{Bae12}. It is seen that the correlation energies $x'$ are close to zero at small $G$ values, which indicate that the values of the Richardson variables are similar to that of the HF solution. Actually, this observation provides a very convenient way to solve the Richardson equations, as is done in the original papers \cite{Rich63} and many recent publications \cite{Bae12}. That is, one start with a vanishing $G$ with known good initial guesses and solve the equations iteratively by slowly increasing its value. This is particularly useful if the ground state of the one of interest.
On the other hand, significant deviations are seen for larger $G$ values where the wave functions get more coherent.
The abrupt changes occur at different $G$ values is related to the fact that the corresponding pairs form a complex conjugate pair starting from those pairing strengths.
In the left panel of the Figure we also done calculations within the HF configuration, i.e., will all single-particle states fully occupied. The pairs can also form complex conjugate partners at certain $G$ values. It indicates that, even for such simple systems, the direct solution of Eq. (1) can be difficult. The solutions for both systems can be found easily by solving Eqs. (\ref{poly3}) and (\ref{poly4}). It may also be useful to mention that in this case the total correlation energy is zero.

\begin{figure}  
\begin{center}
\includegraphics[width=0.45\textwidth]{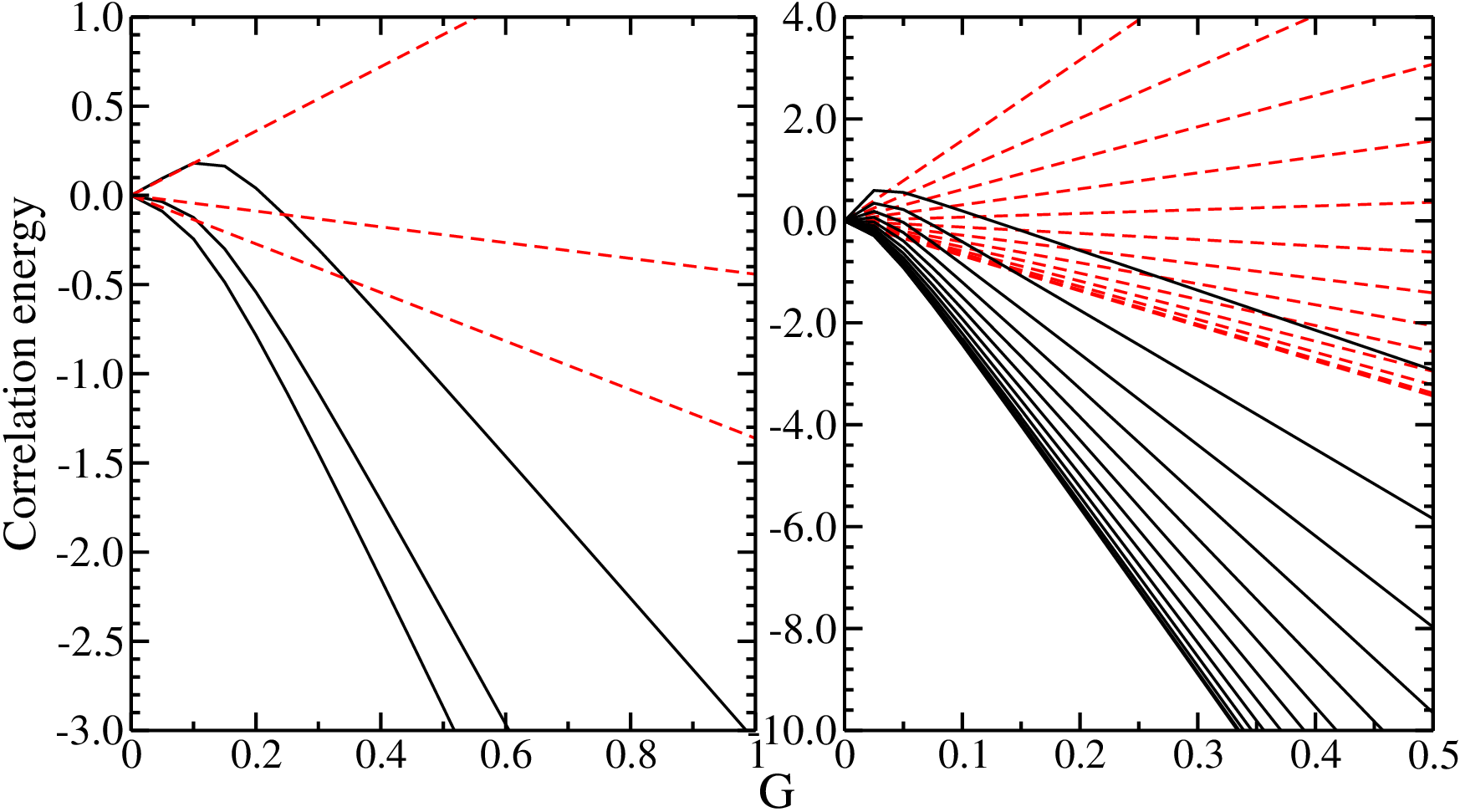}
\end{center}
\caption{\label{two-level} (color online) Correlation energies for each pair (solid line) for two half-occupied systems with $N=6$ (left) and 25 (right) pairs within two $\Omega=6$ (left) and $25$ (right) orbitals separated by one unit. The red dashed lines correspond to those of the fully occupied systems in the single-$j$ orbital.
}
\end{figure}

Systems with large degeneracies can be more challenging to solve since, unlike that of doubly-degenerate systems, one often lacks priori knowledge about the initial guess. To illustrate this point, in Fig. \ref{two-level} we evaluated two half-filled two-level systems with $N=6$ and 25 pairs, respectively. The results are derived by solving Eq. (\ref{poly5}) and compared with those of the fully-filled single-$j$ system.
As can be seen from the figure, for systems within a single-$j$ shell, the Richardson variables all follow a linear relation with $G$ for which one can not apply the iteration approach as mentioned above. There is no abrupt transition in these two cases, which indicate that the solution becomes much easier to solve as long as a good initial guess (i.e., the solution of the  corresponding single-$j$ system) is known. One of the aims of the present work is to find a simple way to solve those single-$j$ systems for which analytic expression for  the total energy has been known.

\begin{figure}  
\begin{center}
\includegraphics[width=0.35\textwidth]{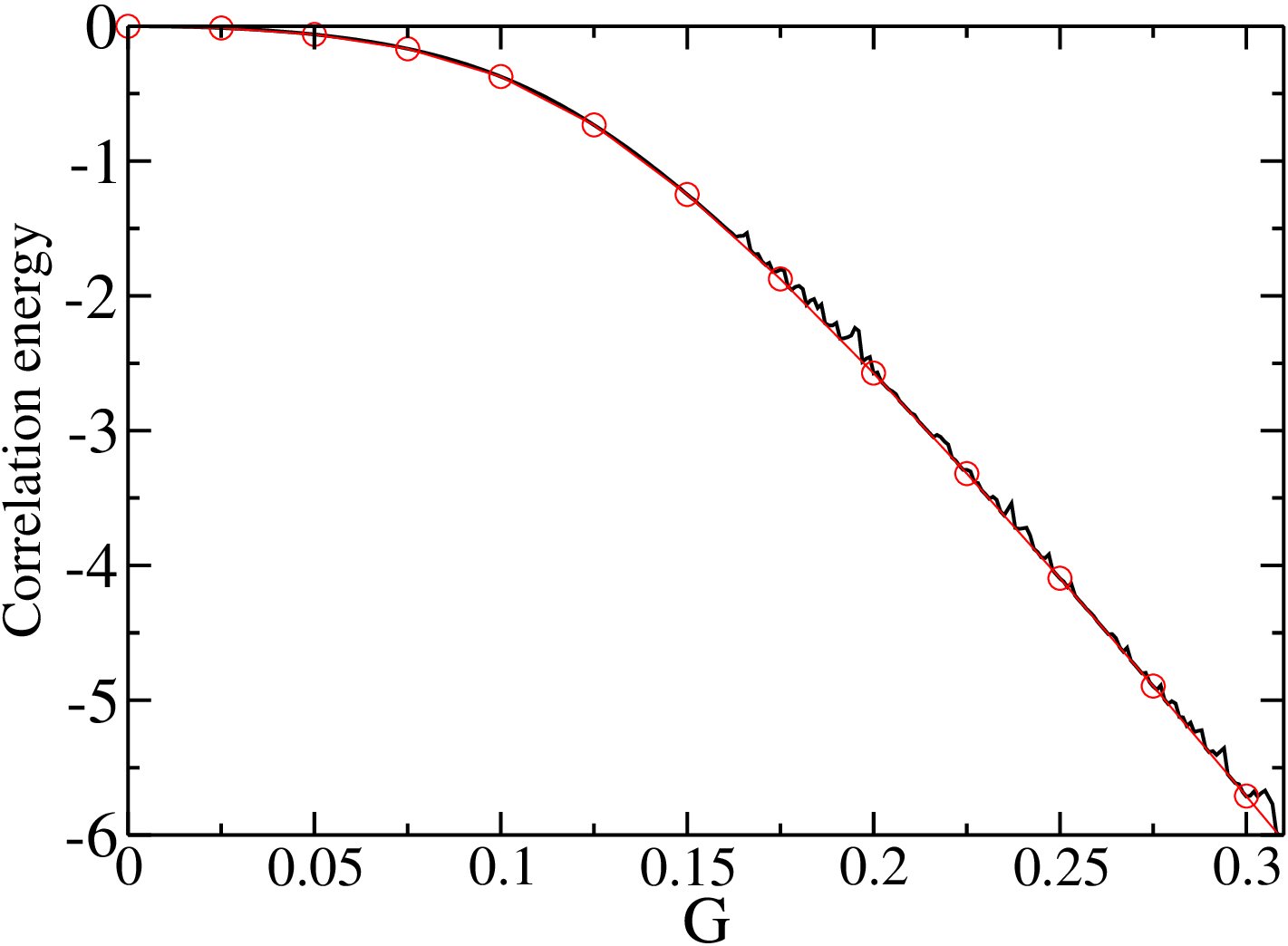}
\end{center}
\caption{\label{fig3} Comparison between the exact (circles) and approximate (black solid line) solutions for total correlation energy of a half-filled system with $N=6$ pairs in two $\Omega=6$ orbitals separated by one unit. The approximate solution is derived by solving Eq. (\ref{poly3}) by slightly breaking the degeneracy of the first orbit by 1\%. 
}
\end{figure}

\begin{figure}  
\begin{center}
\includegraphics[width=0.45\textwidth]{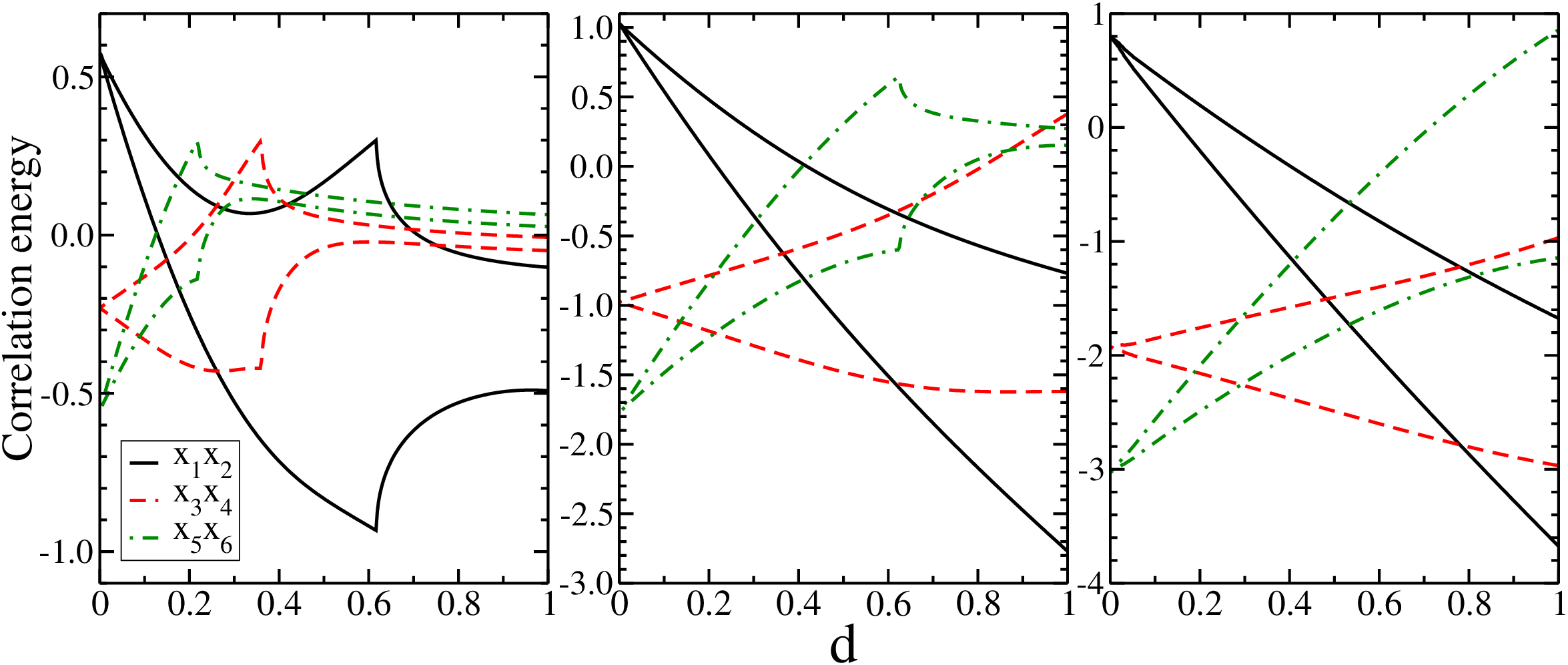}
\end{center}
\caption{\label{fig4} (color online) The correlation energy for each pair for a half-fill doubly degenerate systems with $N=6$ pairs as studied in Fig. \ref{rich} by gradually decreasing the splitting between the filled orbitals. The $G$ values are taken as $G=0.3$, 0.65 and 0.9 for the left, middle, and right panels, respectively.
}
\end{figure}

One interesting thing we notice is that, even though Eq. (\ref{poly3}) and (\ref{poly4}) are ill defined for systems with degeneracies  $\Omega>1$, one can find a good initial guess for such a system by breaking the degeneracy of the system and making it slightly deformed. We analyzed two examples in the present work. Fig. \ref{fig3} corresponds to the half-filled two-level system as studied in the left panel of Fig. \ref{two-level}. The lower orbital is forced to be slightly deformed within the spirit of the Nilsson scheme as  as $\varepsilon_{|j_m|}=\varepsilon_j+d[2|j_m|-(\Omega_j+1)]$ where $m$ denotes the magnetic quantum number and $d=0.01$.
It is seen that, for $G<0.15$, the difference between the exact and approximate solutions are nearly negligible. Small numerical instabilities start to appear at larger $G$ values, which lead to the fluctuations around the exact solution as seen in the figures. A better agreement can be obtained by slightly adjusting the values of $d$ for larger $G$.
Fig. \ref{fig3} indicate that the slightly deformed system can indeed provide a good initial guess for the corresponding system under spherical symmetry. This inspired us to have developed a hybrid algorithm by combining Eq. (\ref{poly3}) and Eq. (\ref{poly5}) which will work for both deformed nuclei as well as spherical and nearly spherical nuclei in the contraction limit.

In Fig. \ref{fig4} we studied the picket-fence model with the orbital separated by $d=1$ as plotted in the right panel of Fig. \ref{rich}. Then we gradually reduce the splitting between the first six filled orbitals to see how the systems involves. The other six levels are kept unchanged. We have chosen three $G$ values: $G=0.3$, 0.65 and 0.9 which have no, two and three complex conjugate pairs, respectively in the original system. In the first two cases, the real Richardson variables can transform to complex ones with decreasing $d$. With $d=1$ the orbitals contract to a single-$j$ shell with $\Omega=6$.

\begin{figure}  
\begin{center}
\includegraphics[width=0.35\textwidth]{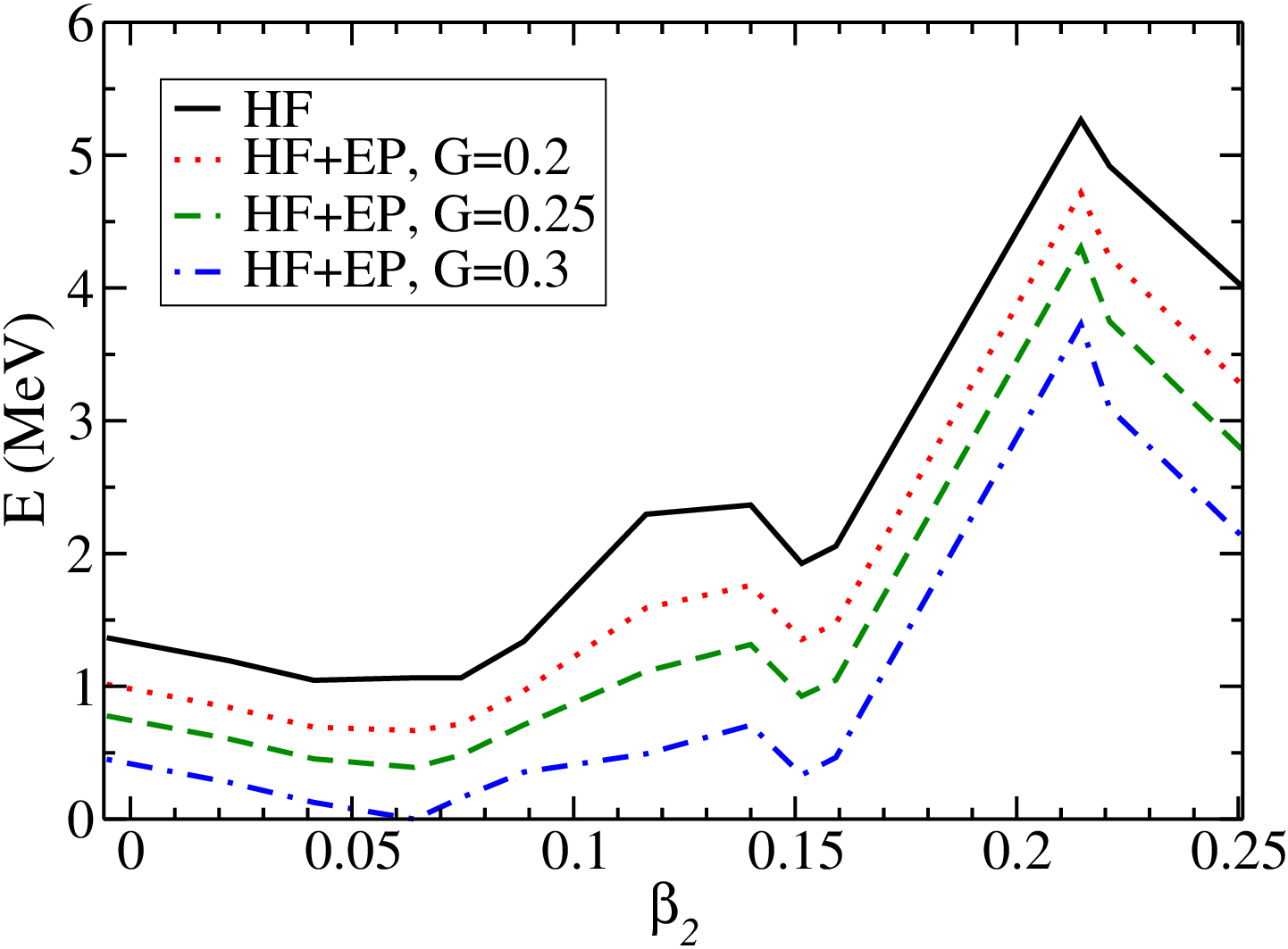}
\end{center}
\caption{\label{fig5} (color online) Relative energies of $^{70}$Ni as a function of quadrupole deformation $\beta_2$ for calculations without pairing (HF) and with pairing with three different strengths. The strengths of the proton and neutron pairings are supposed to be the same for simplicity.
}
\end{figure}

\begin{figure}  
\begin{center}
\includegraphics[width=0.45\textwidth]{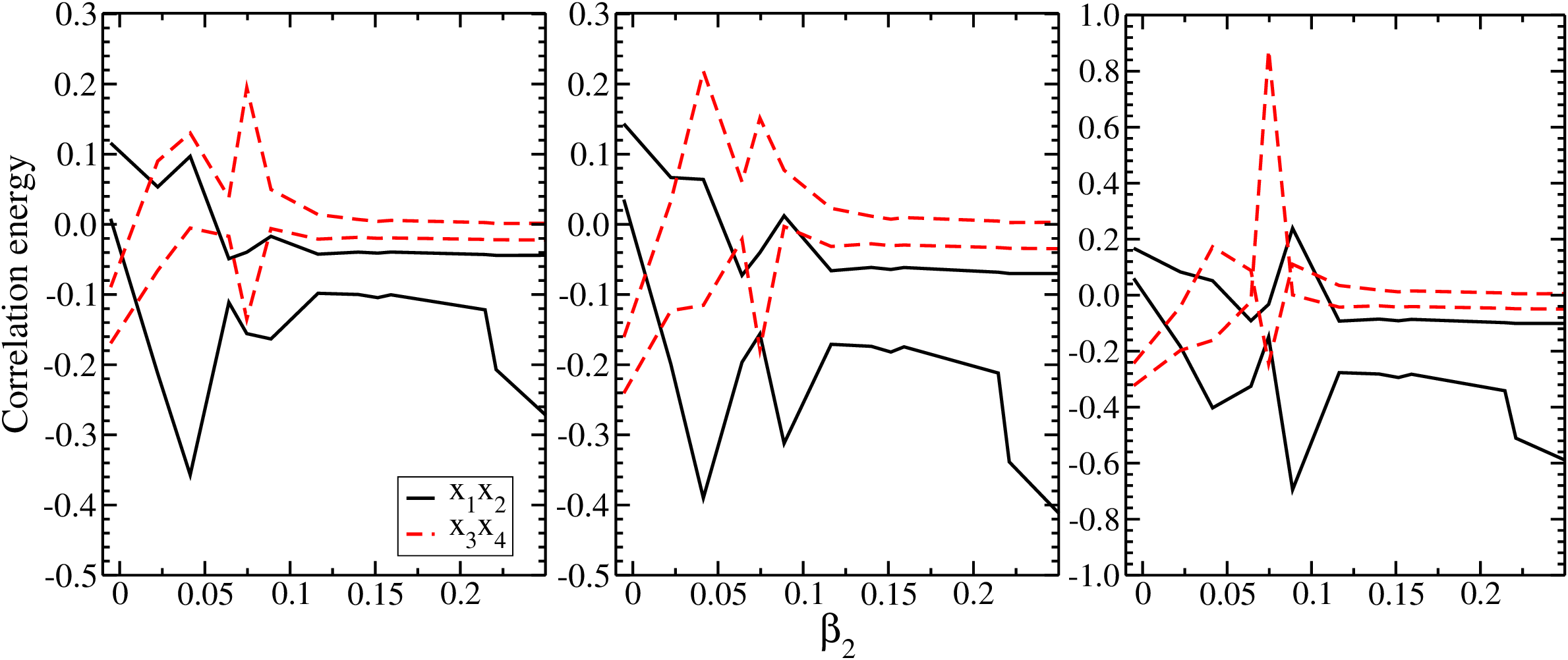}
\end{center}
\caption{\label{fig6} (color online) Correlation energies for the four proton pairs in the $f_{7/2}$ orbital as a function of $\beta_2$ for $G=0.2$ (left), 0.25 (middle) and 0.3 (right).
}
\end{figure}
We then apply the algorithm to study the possible co-existence of different nuclear shapes in neutron-rich Ni isotopes around $^{68}$Ni, which has attracted great attention both theoretically and experimentally \cite{Su2014,Tsu2014} and is supported by our potential energy surface calculations \cite{Wu15,Xu2013247}. As an example,  in Figs. \ref{fig5} \& \ref{fig6} we studied the proton and neutron pairing correlation in the nucleus $^{70}$Ni by applying the exact solution of the pairing on top of the single-particle scheme as determined from standard HF calculations. Only the proton levels between $Z=20$ and 50 and neutron levels between $N=28$ and 56 are considered in solving the pairing Hamiltonian. The single-particle schemes are generated for different deformation by employing 
constrained HF calculations. Calculations are done in the coordinate space with the ev8 code \cite{ev8} and using the SLy4 Skyrme force \cite{Chabanat1998}.
Three different $G$ values are chosen in order to explore its influence on the pair structure. As can be seen from Fig. \ref{fig5}, the ground state of the nucleus $^{70}$Ni favors a nearly spherical shape. In the meanwhile, a low-lying deformed state with $\beta_2$ around 0.15 may be expected. In Fig. \ref{fig6}, we further analyzed the pair structure of the proton pairs and plotted their correlation energies as a function of $\beta_2$. The correlation energies in those three calculations with different $G$ show a similar patten, which may be related to the fact that the $f_{7/2}$ orbital are rather well separated from all other levels due to the presence of $Z=28$ shell closure. But the total correlation energy increase strongly if $G$ increase.

In summary, in this paper we explore further the possibility in solving the non-linear Richardson equation set by using the polynomial approach motivated by recent progress in Refs. \cite{Far11,Guan12}. The equation sets Eq. (\ref{poly3}) and Eq. (\ref{poly4}) can be applied in systems with doubly-degenerate orbitals (or more exactly with $\Omega_{\delta}=1$). We notice that  those two sets can be solved in two possible ways: Either by inserting Eq. (\ref{poly}) into Eq. (\ref{poly3}) directly or through a two-step approach as described in Ref. \cite{Far11}, which involve $N$ (number of pairs) and $M$ (number of orbitals) nonlinear equations, respectively. The former method is easier to solve and can be applied in more general cases. For systems with higher degeneracy, Equation set (\ref{poly5}) is the choice to consider, which can also be reduced to $N$ equations by having derived the expression for the coefficients $b$. We applied the method firstly to a known  picket fence model in Fig. 1 and two two-level models in Fig. 2. 
The drawback of Eq. (\ref{poly5}) is that it contains trivial non-solutions. We found that this may be avoided by taking a good initial guess which may be obtained from the solution of Eq. (\ref{poly3}) by making the system slightly deformed, as is shown in Figs. 3 \& 4. Our combined approach can have great potential in solving the pairing problem for both spherical and deformed nuclei.
As an example, we applied the method to study the energy of the nucleus $^{70}$Ni at different deformation in Fig. 5. Moreover, in Fig. 6 we plotted the corresponding calculated correlation energy for each proton pair in $^{70}$Ni.

\section*{Acknowledgment}
This work was supported by the Swedish Research Council (VR) under grant Nos. 621-2012-3805, and
621-2013-4323.
Part of the
calculations were performed on resources
provided by the Swedish National Infrastructure for Computing (SNIC)
at NSC in Link\"oping and PDC at KTH, Stockholm. CQ also thank X. Guan, Y. Zhang and F. Pan for discussions and their hospitality. Ref. \cite{Far11}  was brought to his attention by J. Dukelsky.


\end{document}